\documentclass[prb,twocolumn,showpacs,preprintnumbers,amsmath,amssymb,floats]{revtex4}

\usepackage{graphicx}

\begin{document}

\title{Normal state electronic structure in the heavily overdoped regime of  Bi$_{1.74}$Pb$_{0.38}$Sr$_{1.88}$CuO$_{6+\delta}$ single-layer cuprate superconductors}
\author{K. Yang$^1$, B. P. Xie$^1$, D. W. Shen$^1$, J. F. Zhao$^1$, H. W. Ou$^1$, J. Wei$^1$, S. Wang$^1$, Y. H. Wang$^1$,
D.H. Lu$^2$, R. H. He$^2$, M. Arita$^3$, S. Qiao$^3$, A. Ino$^4$,
H. Namatame$^3$, M. Taniguchi$^{3,4}$,  F. Q. Xu$^5$, N.
Kaneko$^6$, H. Eisaki$^6$, D.L. Feng$^{1}$}
 \email{dlfeng@fudan.edu.cn}
 \affiliation{$^1$Department of Physics, Applied Surface Physics State Key Laboratory,
  and Synchrotron Radiation Research Center, Fudan University, Shanghai 200433, China}
 \affiliation{$^2$Stanford Synchrotron Radiation Laboratory, Stanford University, Stanford, CA 94305, USA}
 \affiliation{$^3$Hiroshima Synchrotron Radiation Center, Hiroshima University,
Higashi-Hiroshima, Hiroshima 739-8526, Japan.}
  \affiliation{$^4$Graduate School of
Science, Hiroshima University, Higashi-hiroshima, Hiroshima
739-8526, Japan.}
 \affiliation{$^5$National Synchrotron Radiation Laboratory, University of Science \& Technology of China, Hefei 230027, Anhui, China}
 \affiliation{$^6$AIST, 1-1-1 Central 2, Umezono, Tsukuba, Ibaraki, 305-8568 Japan.}
\date{\today}

\begin{abstract}

We explore the electronic structure in the heavily overdoped
regime of the single layer cuprate superconductor
Bi$_{1.74}$Pb$_{0.38}$Sr$_{1.88}$CuO$_{6+\delta}$. We found that
the nodal quasiparticle behavior is dominated mostly by phonons,
while the antinodal quasiparticle lineshape is dominated by spin
fluctuations. Moreover, while long range spin fluctuations
diminish at very high doping, the local magnetic fluctuations
still dominate the quasiparticle dispersion, and the system
exhibits a strange metal behavior in the entire overdoped regime.

\end{abstract}


\maketitle

The high-$T_c$ cuprate superconductors (HTSC's) have been studied
for almost 20 years.\cite{Millisreview} Anomalous and complicated
phenomenology in the underdoped regime of HTSC's such as
pseudogap,\cite{TimuskPGreview} resonant modes in spin fluctuation
spectra\cite{neutronreview} and possible charge
ordering,\cite{DavisCDW,Tranquada} have led to a broad spectrum of
theoretical proposals from the very conventional BCS-type models
\cite{Levin} to the very exotic ones involving various gauge
fields.\cite{Leewenreview04} Recent angle resolved photoemission
spectroscopy (ARPES) experiments resolved kinks of the
quasiparticle dispersions in both nodal and antinodal region of
the Brillouin
zone.\cite{ARPESHTSC,Alenature,Kaminski01,mag,Adam,Cuk} Whether
these kinks are originated from electron-phonon or electron-magnon
interaction is currently
debated.\cite{Alenature,Kaminski01,mag,Adam,Cuk,TD} Nevertheless,
the role of phonon in high temperature superconductivity is being
revisited.\cite{Nagaosa}

To date, most experimental studies have been focusing on the
anomalous, and certainly interesting, underdoped regime. On the
other hand, although to be validated by experiments, the overdoped
half of the phase diagram, particularly the heavily overdoped
regime, is considered to be a ``normal" metal regime where
correlations are negligible. The electronic structure in this
regime was seldom studied.\cite{Yusof,LSCO,sato1,sato2} To build a
comprehensive picture of high temperature superconductivity, it is
crucial to study the properties of the entire overdoped regime,
particularly the heavily overdoped regime, where little data is
available at this stage. By examining the overall picture, one
hopefully can understand the complications in the HTSC
phenomenology better and resolve some of the current
controversies, and thus sort out properties that are essential for
the high temperature superconductivity.

In this paper, we report a systematic ARPES study of the
electronic structure in the heavily overdoped regime of HTSC's up
to the extreme doping level where the superconductivity vanishes
(referred as ultra-overdoped regime hereafter). Our data indicate
that lattice effect dominates the nodal quasiparticle scattering,
while the spin fluctuations dominate the antinodal region.
Although long range spin fluctuations diminish at high doping
together with the weakening superconductivity, the local
antiferromagnetic fluctuations still dominate the quasiparticle
dispersion, and correlations are still strong.

High quality superstructure-free
Bi$_{1.74}$Pb$_{0.38}$Sr$_{1.88}$CuO$_{6+\delta}$ (Pb-Bi2201)
single crystal was grown by floating-zone technique. Unlike
Bi$_2$Sr$_2$CaCu$_2$O$_{8+\delta}$ (Bi2212), Pb-Bi2201 is a
single-layer cuprate and can be doped into much higher doping
level. Through Oxygen or Argon annealing, we have prepared
overdoped samples with superconducting phase transition
temperature $T_c$'s (dopings) of 0K (0.27), 5K (0.26), 7K (0.258),
10K (0.252), 16K (0.24) and 22K (0.225), where the hole
concentration $x$'s are estimated based on the empirical formula
$T_c=T_{c,opt}[1-82.6(x-0.16)^2]$,\cite{dopingformula} and
$T_{c,opt}=34$K for Bi2201. For comparison, we also studied a
Pb-free optimally doped Bi2201 ($T_c=34K$, $x=0.16$). Transition
widths of these samples are typically $1\sim 2$K. Moreover, unlike
La$_{2-x}$Sr$_x$CuO$_{4}$ (LSCO), the cleavage surface of Bi2201
is non-polar and stable, suitable for angle resolved photoemission
studies.  ARPES experiments were performed at the Beamline 5-4 of
Stanford Synchrotron Radiation Laboratory (SSRL), and Beamline 9
of Hiroshima Synchrotron Radiation Center (HiSOR). Both beamlines
are equipped with a Scienta SES200 electron analyzer, with a
typical angular resolution of $0.3^\circ$ and an energy resolution
of 10\,meV. If not specified, the data were taken at $10\sim 15$K
above $T_c$ with 22.7eV photons.

\begin{figure}[t!]
\includegraphics[width=8.5cm]{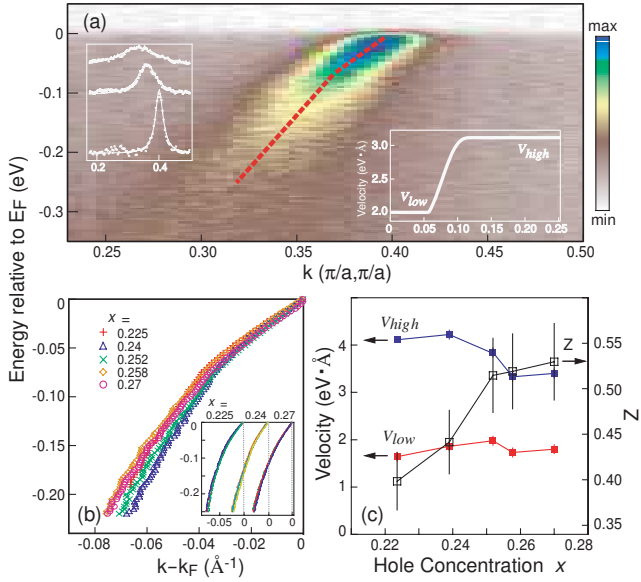}
\caption{(a) Normal state photoemission intensity as a function of
energy and nodal momentum for $x=0.26$. The inset on the left is
MDC's at different binding energies that can be fit well with a
Lorentzian. The dispersion is extracted through the MDC analysis and
shown as the thick dashed line. The inset on the right is the band
velocity based on the dispersion, where it is a fixed value at low
or high binding energies defined as $V_{low}$ or $V_{high}$.
(b)Dispersion extracted from MDC analysis in the nodal direction for
various doping levels. Inset: Dispersions for $x=0.225$ and T=30,
100, 200K; $x=0.24$ and T=60, 120, 190, 250K; and $x=0.27$ and T=10,
100, 150, 200, 250K. (c) $V_{low}$, $V_{high}$ and
$Z\equiv$$V_{low}$/$V_{high}$ as a function of doping.}
 \label{Z}
\end{figure}

\begin{figure}[t!]
\includegraphics[width=8.5cm]{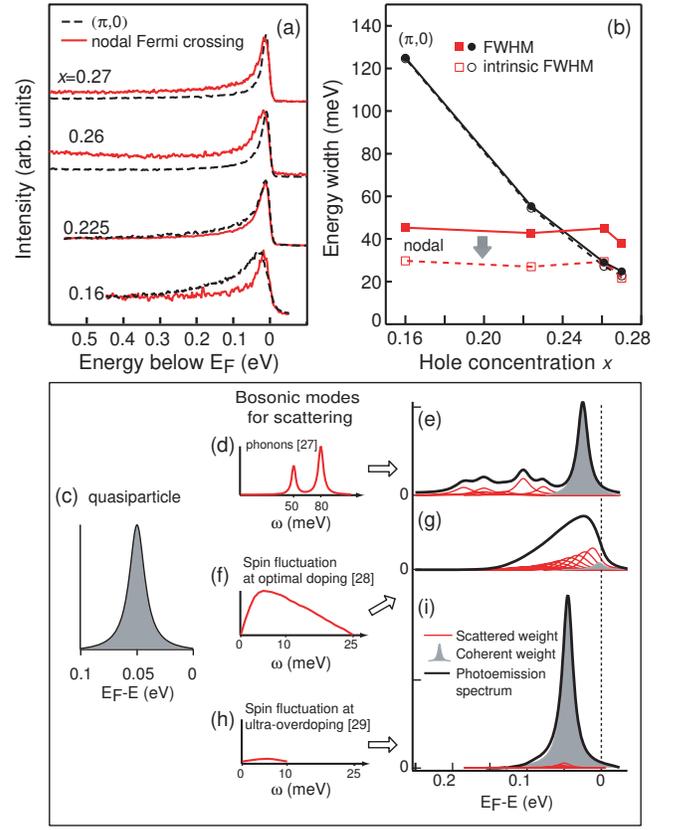}
 \caption{(a)The
evolution of photoemission spectral lineshape of the Bi2201 system
as a function of doping. Compared with the spectrum at nodal Fermi
crossing (solid curves), the $(\pi,0)$ (dashed curves) gradually
sharpens up with increased hole doping. (b) FWHM of the spectra in
(a) are displayed in solid symbols, and the intrinsic FWHM after
deducting the resolution effects are shown in empty symbols. (c-i)
Numerical simulations of the effects of quasiparticle-bosonic mode
interactions on photoemission lineshape. When the original
quasiparticle (c) is scattered multiple times by discrete optical
phonon modes (d), or spin fluctuations at optimal doping (f), or at
ultra-overdoping (h), the resulting photoemission lineshape are
illustrated in (e), (g), and (i) respectively. Realistic parameters
are implemented in the simulation. The intrinsic quasiparticle width
is set to 20meV; the multiple optical phonon modes in the relevant
energy region are based on maximum entropy analysis results on
Bi2201 spectra in Ref.\cite{Non}; and the spin fluctuation spectra
near $(\pi,\pi)$ are taken from inelastic neutron scattering
measurements in the normal state for optimally-doped,\cite{Aeppli}
and ultra overdoped $T_c\sim4.5K$ single layer LSCO
systems,\cite{Wakimoto04} since the data on Bi2201 are unavailable.
The residual quasiparticle weight $Z$ in (e) is set to 0.5, which
restrains the scattering strength. The scattering strength in (g) is
chosen so that the spectral linewidth is close to the experimental
one after deducting a background. The relative strength between spin
fluctuation intensities in (f) and (h) are based on an extrapolation
of the doping dependence of maximum $\chi^"(\omega)$ in
Ref.\cite{Wakimoto04}, which in turn determines the scattered weight
in (i).} \label{dope}
\end{figure}

We start with the examination of the nodal quasiparticle behavior.
Fig.\,\ref{Z}a shows the nodal photoemission intensity map, where
a kink in dispersion is clearly visible even for the $x=0.26$ high
doping level.\cite{Alenature} As illustrated by the thick dashed
line, one can extract the effective band velocities $V_{high}$ and
$V_{low}$ for dispersion below and above the kink energy scale
respectively (inset of Fig.\,\ref{Z}a). Dispersions and effective
band velocities for various dopings are displayed in
Figs.\,\ref{Z}b and \ref{Z}c respectively. Similar to the LSCO
system,\cite{Zhounature} $V_{low}$ here is also almost doping
independent, while $V_{high}$ decreases slowly with increased
doping. Recently, it has been shown that the coherent
quasiparticle spectral weight $Z$ can be estimated as
$V_{low}/V_{high}$ for cuprates,\cite{Randeria} which increases
from $0.4$ at $x=0.225$, to $0.53$ at $x=0.27$(see
Fig.\,\ref{Z}c). These agree quite well with the calculations by
Paramekanti \textit{et al.}\cite{RanderiaPRL} At different
temperatures (inset in Fig.\,\ref{Z}b), the dispersion varies very
little, indicating that $Z$ does not decrease with rising
temperature. Therefore, in the highly overdoped regime of Bi2201,
we do not observe a coherent-incoherent transition with increasing
temperature as a recent study of overdoped Bi2212
claimed.\cite{Kaminsikitransition}

Fig.\,\ref{dope}a compares the photoemission spectrum at nodal
Fermi crossing and that at $(\pi,0)$ for four different dopings
ranging from the optimal doping level to ultra-overdoped $T_c=$0K
sample. Remarkably, the $(\pi,0)$ spectrum sharpens up
dramatically with increased doping, while the nodal spectral
linewidth is quite doping independent.  At very high doping, the
$(\pi,0)$ spectrum appears to be even sharper than the nodal
spectrum. The full-width-half-maximum (FWHM) of these spectra are
plotted in Fig.\,\ref{dope}b as solid symbols. After removing the
energy broadening from finite energy resolution and angular
resolution, which is significant in nodal direction due to the
fast dispersion,\cite{knote} one obtains the intrinsic energy
width shown as open symbols. The scattering rate at $(\pi,0)$
almost decreases by a factor of five in a linear fashion, while
the scattering of the nodal quasiparticles remains roughly
constant. In the ultra-overdoped regime, the nodal and $(\pi,0)$
quasiparticles have the same linewidth, \textit{i.e.}, the
scattering across the Fermi surface becomes isotropic.

Based on the above experimental observations, one can retrieve
information on the interactions between electrons and bosons in
the nodal and antinodal regions. In cuprates, the possible
candidates for the bosons are phonons and spin excitations. For
the nodal quasiparticles, the linewidth is almost doping
independent, while $Z$ increases with doping. This peculiar
behavior resembles the photoemission spectrum of Hydrogen
molecules, where the vibration modes take the quasiparticle
spectral weight from the main peak and redistribute it in multiple
mode energies away without affecting the main peak
linewidth.\cite{H2,ARPESHTSC} In solid, as simulated in Fig.2d and
2e, it is only when electrons interact with \textit{discrete}
bosonic modes with frequencies larger than the original
quasiparticle width, could the quasiparticle keep a fixed width
while its spectral weight increases with weakening coupling
strength. Half-breathing phonon maybe is a candidate to explain
our data, it was suggested to interact strongly with the nodal
quasiparticles in Bi2212 system.\cite{TD,Alenature} Moreover,
since phonon spectrum does not vary significantly with doping, the
increasing screening of electron-phonon interactions can explain
the moderate 30\% increase of  $Z$  from $x=0.225$ to $0.27$. On
the other hand, we found that in Fig.2f-i, because the spin
excitation is a broad spectrum starting from zero frequency in the
normal state,\cite{Aeppli,Wakimoto04} its scattering with
electrons causes the resulting spectral linewidth a large doping
dependence. Furthermore, spin excitation spectral weight drops
rapidly to a negligible level at high doping,\cite{Wakimoto04}
which is hard to count for the loss of 47\% (\textit{i.e.} $1-Z$)
of the nodal coherent quasiparticle weight for $x=0.27$.
Therefore, These observations indicate that optical phonons may
dominate the scattering of quasiparticles in the nodal region.

\begin{figure}[t!]
\includegraphics[width=8.5cm]{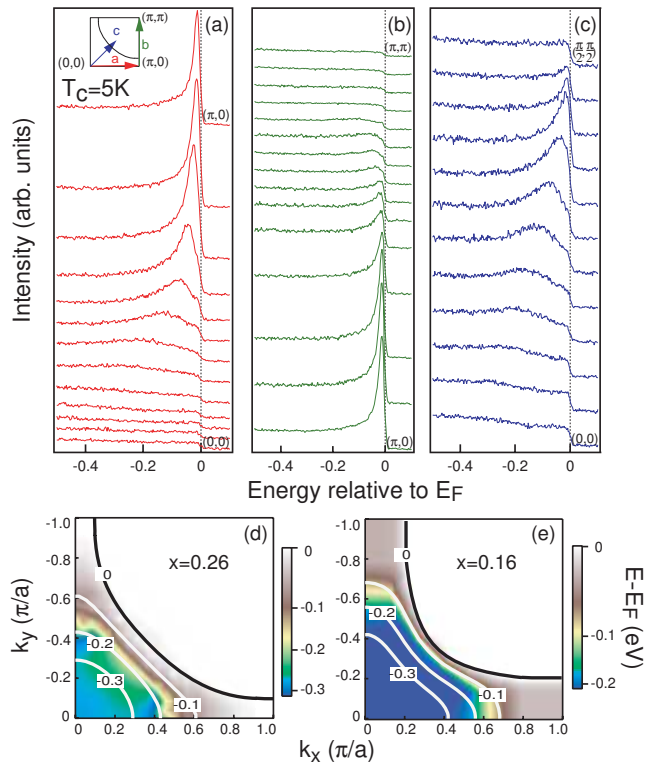}
\caption{(a-c) Photoemission spectra along three high-symmetry lines
as indicated in the inset of panel (a) for $x=0.26$. Data is taken
at 12K with a Helium lamp. (d,e)Image plots of dispersions extracted
from the photoemission peak position of $x=0.26$ and $x=0.16$
respectively. The contour curves are the tight-binding fits to the
dispersions.}
 \label{EDC}
\end{figure}

Contrary to the nodal case, the \textit{continuously} fast
decreasing $(\pi,0)$ linewidth with doping (Fig.2b) can be
explained by the simulation in Fig.2f-i. The broad $(\pi,0)$
spectrum at optimal doping can be attributed to the strong
scattering from the antiferromagnetic fluctuations near
$(\pi,\pi)$ (Fig.2g).\cite{Aeppli} Consistently, when spin
fluctuations are significantly weaker in the ultra-overdoped
regime as shown by recent neutron scattering data
(Fig.2h),\cite{Wakimoto04} spectral linewidth gradually recovers
its original quasiparticles width (Fig.2i), eventually resulting
in an isotropic linewidth across the Fermi surface for $T_c\sim
0K$. Therefore, the doping and momentum dependence of the
quasiparticle lifetime indicate the dominant role of spin
fluctuations to the antinodal quasiparticles. As for phonon
effects in this region, the out-of-plane $B_1g$ buckling phonon
was suggested to interact strongly with the antinodal
quasiparticles in bilayer systems due to the nonzero crystal field
along $c$-axis at the $CuO_2$ planes.\cite{TD95,Cuk,TD} However,
for single layer Bi2201, the surroundings of the $CuO_2$ plane are
symmetric, thus the crystal field induced electron-buckling phonon
coupling is negligible in Bi2201.\cite{TD95} For the in-plane
half-breathing phonon, it has been suggested that it only
interacts weakly with the electrons in the antinodal
region.\cite{TD} However, we note that at ultra-overdoping, when
spin fluctuations diminish, effects of other phonons might become
relatively pronounced in the quasiparticle lineshape (not included
in Fig.2i for simplicity).

Having studied the nodal and $(\pi,0)$ quasiparticles, we now turn
to the overall dispersion. Fig.\,\ref{EDC}(a-c) present the normal
state ARPES spectra measured on the ultra-overdoped $x=0.26$
sample along the high symmetry directions of the first Brillouin
zone. Tracking the peak position of the spectrum, the
quasiparticle dispersion is plotted in false color in
Fig.\,\ref{EDC}(d) for $x=0.26$ and \ref{EDC}(e) for $x=0.16$ for
comparison, which shows that the overall dispersion energy scale
is not very sensitive to doping. Moreover, we observed hole-like
Fermi surface even at the highest doping level. The data can be
well fit with an effective tight-binding model (shown as contour),
 $\varepsilon(k)=e_0-2t(\cos k_x+\cos k_y )
-4t'\cos k_x \cos k_y - 2 t''(\cos 2k_x+ \cos 2k_y)$
 .\cite{NormanPRB,Liechenstein} We
got the effective hopping parameters: \\
 \centerline{
  \begin{tabular}{c|cccc}
 $x$  &    $t$ (eV) & $t'$ (eV)  & $t''$ (eV)   &  $e_0$ (eV) \\
   \hline
   \hline
   0.26 &  $0.164(2) $   & $-0.035(1)$&$0.015(1) $   &$0.182(1) $
   \\
   0.16 &  $0.130(5)$  &   $-0.010(3)$ &   $0.034(2)$ & $0.062(2)$ \\
  \end{tabular}
}
 \noindent If the correlations at the ultra-overdoped regime were
very weak, one would expect the effective hopping between nearest
neighbors, $t$, to be much larger, close to 0.4eV as estimated by
band structure calculations. But the small $t$ here clearly
indicates that the dispersion energy scale is still of the
exchange energy $J$. It is interesting to notice that although the
long range magnetic fluctuations are much reduced when approaching
the ultra-overdoped regime, the local antiferromagnetic
correlations still determine the hopping behavior of the holes.
This is consistent with the recent numerical studies based on the
$t$\,-\,$t'$\,-\,$t''$\,-\,$J$ model, showing that the dispersion
of the centroid of the occupied state has very small doping
dependency.\cite{Tohyama}

\begin{figure}[t!]
\includegraphics[width=8.5cm]{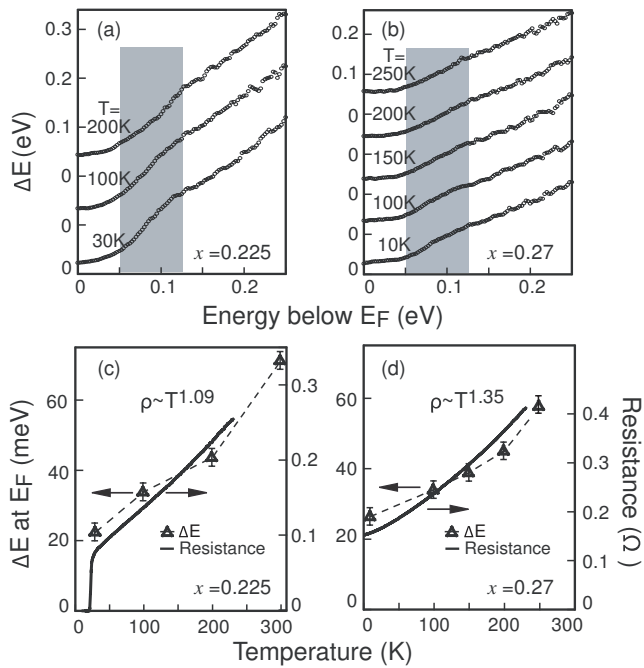}
\caption{(a,b)Binding energy dependence of the quasiparticle inverse
lifetime along the nodal direction for $x=0.225$, and $0.27$
respectively. The shaded regions indicate kink effects in
scattering. (c,d)Temperature dependence of the nodal quasiparticle
inverse lifetime at $E_F$ and resistance for $x=0.225$, and $0.27$
respectively.
 }
 \label{NFL}
\end{figure}

The highly overdoped system possesses both strong correlation and
well defined quasiparticle. The energy dependence of the nodal
quasiparticle scattering rate, \textit{i.e.} $\Delta E\equiv
V\times\Delta k$ is plotted in Fig.\,\ref{NFL}a,b for two dopings,
where $V$ is the band velocity indicated in the inset of
Fig.\,\ref{Z}a. The shaded regions correspond to the kink, which
gives an upturn in scattering rate due to electron-boson
interactions, where these bosons are most likely phonons as
discussed above. The low energy region can be fit with a power law
function, $\Delta E \sim \omega^{\alpha}$, but $\alpha$ varies from
1 to 3, depending on the choice of the fitting energy window.
Meanwhile, the high energy portion above the kink region is always
linear. As a function of temperature, the in-plane resistance is
plotted in Fig.\,\ref{NFL}c,d together with  the nodal quasiparticle
scattering rate at Fermi energy. Close-to-linear behavior is
observed for
 $x=0.225$, while for $x=0.27$, a superlinear temperature
dependence is observed, $\rho\sim T^{1.35}$. Similarly, early
resistivity studies of ultra-overdoped Tl$_2$Ba$_2$CuO$_{6+\delta}$
have shown superlinear behavior as well.\cite{Taillefer} These
temperature dependence behavior and the interesting energy
dependence indicate that the system is still ``strange metal" state
in the ultra-overdoped regime.

The non-Fermi liquid behavior near the optimal doping was related
to a possible quantum critical point in the
vicinity.\cite{Valla,Marel,QCP} The transition from a marginal
Fermi liquid behavior toward a more Fermi-liquid-like behavior
with tuning of the doping does resemble the quantum critical
phenomena observed in Sr$_3$Ru$_2$O$_7$ and
YbRh$_2$Si$_2$\cite{SRuO,YRS}. Moreover, in various doping levels
and temperatures that were sampled, we so far do not observe any
abrupt change in the electronic structure that might be related to
a first order transition.  Whether or not there is one or more
quantum critical points still needs more decisive evidence.

The nodal quasiparticles display many doping-insensitive
properties such as the linewidth, $v_{low}$. We find them not
affected  by the long range magnetic fluctuations. Since nodal
quasiparticles are responsible for many transport properties, our
observations put strong constraints to theories of transport. For
the antinodal region where the gap is the largest, the data
suggest that long range magnetic fluctuations play an important
role. The fact that $d$-wave superconductivity vanishes when the
quasiparticle inverse lifetime across the Fermi surface becomes
isotropic is very intriguing. Moreover, the quasiparticle
dispersion is renormalize by roughly a factor of $x$ from the bare
band, indicative of that the local magnetic correlations are still
surprisingly strong up to the ultra-overdoped
regime.\cite{anderson}

We conclude by pointing out that the ultra-overdoped regime
exhibits many unconventional behaviors that may be fundamentally
related to the mechanism of high temperature superconductivity,
which calls for more comprehensive theory that could explain the
phenomenology over the entire doping range.

{\it Acknowledgements:} DLF would like to thank Profs. Q. H. Wang,
J. X. Li and J. P. Hu for helpful discussions. This work is
supported by NSFC, Shanghai Municipal government, and Japan-China
core university program; SSRL is operated by the DOE Office of
BES, Divisions of Chemical Sciences and Material Sciences.


\vspace*{-0.6 cm}
\begin{references}

\bibitem{Millisreview}J. Orenstein and A. J. Millis, Science {\bf 288},
468 (2000).
\bibitem{TimuskPGreview}T. Timusk, B. Statt, Rep. Prog. Phys. {\bf
62}, 61 (1999)
\bibitem{neutronreview}Y. Sidis {\sl et al.}, Phys. Status Solidi
B {\bf 241}, 1204 (2004).
\bibitem{DavisCDW} T. Hanaguri {\sl et al.} Nature (London) {\bf 430}, 1001 (2004).
\bibitem{Tranquada}J. M. Tranquada {\sl et al.} Nature (London) {\bf 375}, 561 (1995).
\bibitem{Levin}Q. Chen, I. Kosztin, B. Jank$\acute{o}$, and K.
Levin, Phys. Rev. B \textbf{59}, 7083 (1999).
\bibitem{Leewenreview04}P. A. Lee, N. Nagaosa, and X.-G. Wen,
arXiv:cond-mat/0410445 (2004)
\bibitem{ARPESHTSC}A. Damascelli, Z. Hussain, and Z.-X. Shen, Rev. Mod.
Phys. {\bf 75}, 473 (2003).
\bibitem{Alenature}A. Lanzara {\sl et al.}, Nature (London) {\bf 412}, 510
(2001).
\bibitem{Kaminski01}A. Kaminski \textit{et
al.}, Phys. Rev. Lett. \textbf{86}, 1070 (2001).
\bibitem{mag}P. D. Johnson \textit{et al.}, Phys. Rev. Lett. \textbf{87},
177007 (2001).
\bibitem{Adam} A. D. Gromko \textit{et al.}, Phys. Rev. B,
68, 174520 (2003).
\bibitem{Cuk}T. Cuk {\sl et al.}, Phys. Rev. Lett. {\bf 93}, 117003 (2004).
\bibitem{TD}T. P. Devereaux {\sl et al.}, Phys. Rev. Lett. {\bf 93}, 117004 (2004).
\bibitem{Nagaosa}Z. X. Shen {\sl et al.}, Philos. Mag. B {\bf 82}, 1349 (2002).
\bibitem{Yusof}Z. M. Yusof \textit{et al.}, Phys. Rev. Lett. \textbf{88},
167006 (2002).
\bibitem{LSCO}A. Ino \textit{et al.}, Phys. Rev. Lett. \textbf{81},
2124 (1998).
\bibitem{sato1}T. Sato \textit{et al.}, Physica C \textbf{341}, 2091 (2000)
\bibitem{sato2}T. Sato \textit{et al.}, Physica C \textbf{364},
590 (2001)
\bibitem{dopingformula}M.R. Presland {\sl et al.}, Physica \textbf{C 176}, 95 (1991).
\bibitem{Zhounature}X. J. Zhou {\sl et al.}, Nature (London)
{\bf 423}, 398 (2003).
\bibitem{Randeria}M. Randeria, A. Paramekanti, and N. Trivedi, Phys. Rev. B {\bf 69}, 144509
(2004).
\bibitem{RanderiaPRL}A. Paramekanti, M.
Randeria, and N. Trivedi, Phys. Rev. Lett. {\bf 87}, 217002
(2001).
\bibitem{Kaminsikitransition}A. Kaminski  \textit{et al.}, Phys. Rev. Lett. \textbf{90},
207003 (2003).
\bibitem{knote} Linewidth broadening of the finite momentum resolution is in the order of $\Delta k \times v_F$, where
$\Delta k$ is the momentum resolution and $v_F$ is Fermi velocity.
\bibitem{H2}D. W. Turner, A. D. Baker, C. Baker, and C. R.
Brundle, \textit{Molecular Photoelectron Spectroscopy}, (Wiley,
New York, 1970).
\bibitem{Non}W. Meevasana {\sl et al.}, preprint.
\bibitem{Aeppli}G. Aeppli {\sl et al.}, Science {\bf 278}, 1432 (1997).
\bibitem{Wakimoto04}S.Wakimoto {\sl et al.}, Phys. Rev. Lett. {\bf
92}, 217004 (2004).
\bibitem{TD95}T. P. Devereaux {\sl et al.}, Phys. Rev. B {\bf 51},
505 (1995).
\bibitem{NormanPRB}M. R. Norman {\sl et al.}, Phys. Rev. B {\bf
52}, 615 (1995).
\bibitem{Liechenstein}A. I. Liechtenstein {\sl et al.}, Phys. Rev. B {\bf
54}, 12505 (1996).
\bibitem{Tohyama}T. Tohyama, Phys. Rev. B {\bf 70}, 174517 (2004).
\bibitem{Taillefer}A. P. Mackenzie {\sl et al.}, Phys. Rev. Lett. {\bf 71}, 1238
(1993); C. Proust {\sl et al.}, {\sl ibid.} {\bf 89}, 147003
(2002).
\bibitem{Valla}T. Valla {\sl et al.}, Science {\bf 285}, 2110
(1999).
\bibitem{Marel}D. van der Marel {\sl et al.}, Nature (London) {\bf
425}, 271 (2003).
\bibitem{QCP}S. Sachdev, Science {\bf 288}, 475 (2000).
\bibitem{SRuO}S. A. Grigera \textit{et al.}, Science \textbf{294}, 329 (2001).
\bibitem{YRS}J. Custers \textit{et al.}, Nature \textbf{424}, 524 (2003).
\bibitem{anderson}P. W. Anderson {\sl et al.}, J. Phys. Cond. Mat. {\bf 16}, R755
(2004); F. C. Zhang \textit{et al.}, Supercond. Sci. Tech.
\textbf{1}, 36 (1988); Q. H. Wang \textit{et al.},
cond-mat/0506712.
\end{references}
\end{document}